\documentclass[12pt]{iopart}

\usepackage{graphicx}

\begin{document}

\title[Self-assembled hexagonal double fishnets as NIMs]{Self-assembled hexagonal double fishnets as negative index materials}

\author{Kristof Lodewijks$^{1,2,*}$, Niels Verellen$^{1,2,3}$, Willem Van Roy$^1$, Gustaaf Borghs$^1$ and Pol Van Dorpe$^{1,**}$}

\address{$^1$ Interuniversity Microelectronics Center (IMEC), Kapeldreef 75, B-3001 Leuven, Belgium}
\address{$^2$ KU Leuven departement elektrotechniek ESAT, Kasteelpark arenberg 1, B-2001 Leuven, Belgium}
\address{$^3$ INPAC-Institute for Nanoscale Physics and Chemistry, Nanoscale Superconductivity and Magnetism and Pulsed Fields Group, K. U. Leuven Celestijnenlaan 200 D, B-3001 Leuven, Belgium}
\ead{$^{*}$\mailto{kristof.lodewijks@imec.be}, $^{**}$\mailto{ pol.vandorpe@imec.be }}

\begin{abstract}
We show experimentally the successful use of colloidal lithography for the fabrication of negative index metamaterials in the near-infrared wavelength range. In particular, we investigated a specific implementation of the widely studied double fishnet metamaterials, consisting of a gold-silica-gold layer stack perforated by a hexagonal array of round holes. Tuning of the hole diameter allows us to tailor these self-assembled metamaterials both as single- ($\epsilon < 0$) and double ($\epsilon < 0$ and $\mu < 0$) negative  metamaterials.
\end{abstract}

\maketitle

\section{Introduction}
Since the introduction of negative index materials (NIMs) by Veselago \cite{Veselago} and the discovery of the possibility of realizing sub-wavelength resolution for imaging devices based on these metamaterials by Pendry \cite{Pendry}, many different designs have been proposed to make the "perfect lens" dream reality. A notable and widely studied geometry in the visible and near-infrared wavelength range is the double fishnet structure \cite{Dolling1,Dolling2,Dolling3,Chettiar,Xiao}, which consists of a stack of metal-insulator-metal (MIM) layers perforated by a periodic array of rectangular holes. The pioneering work by Dolling et al. \cite{Dolling2} demonstrated the reversal of the phase velocity in the double fishnet metamaterials, and more recently, also negative refraction was observed in a multilayer fishnet prism structure \cite{Valentine}. The negative refractive index behavior of these metamaterials is governed by a magnetic resonance that is excited in the MIM cavities in between the holes and the negative permittivity of the metal layers \cite{Mary}. At the magnetic resonance frequency, gap plasmons are excited on the top- and bottom interface of the insulator layer of the MIM cavities, which give rise to a strong magnetic dipole that is out of phase with the magnetic field component of the incident wave. As a result, the effective magnetic permeability $\mu$ is strongly lowered at this resonance and can even reach negative values. In combination with the negative permittivity $\epsilon$ of the metal layers in this wavelength range, this gives rise to a negative value for the refractive index. Based on the real parts (') of the effective material parameters of the metamaterial, NIMs can be classified as single negative metamaterials (SN-NIMs with $\epsilon < 0$ and $\mu > 0$ while $n < 0$) or double negative metamaterials (DN-NIMs with $\epsilon < 0$ and $\mu < 0$ while $n < 0$). The 
 of merit (FOM) for the magnetic resonance in NIMs is defined as the amplitude of the ratio between the real part (') and the imaginary part (") of the refractive index ($FOM = | n' / n" |$). Depending on the single- or double negative nature of the metamaterial, low (SN-NIMs) or high (DN-NIMs) values of the FOM are observed. As absorption in the metal layers becomes more and more important when getting closer to the plasma frequency \cite{Zhou,Klein}, it is important to optimize the metamaterial design in order to create a strong magnetic resonance. This is the main reason why the double fishnet structures obtain better FOMs than U-shaped Split-Ring Resonator (SRR) \cite{Klein} based structures, as the magnetic resonance cavity can be seen as a SRR with two instead of one slit, which allows to increase the saturation frequency of the magnetic resonance \cite{Zhou}. A major drawback of the structures reported to date is that the fabrication involves expensive and low-throughput lithography steps such as e-beam lithography or focussed ion beam (FIB) milling, which limits the potential for usage of these metamaterials in large-scale applications.

In this letter, we show that we can circumvent this limitation by using nanosphere lithography, which allows us to create large-area double fishnet metamaterials consisting of a MIM layer stack perforated by a hexagonal array of holes. The hole pattern is generated using a self-assembled close packed monolayer of $550$ nm polystyrene (PS) beads, which is deposited by spin coating. By varying the bead diameter, we can change the pitch of the bead array, while the hole size can be tuned by the bead shrinking process, which allows us to create both single- and double negative metamaterials. Moreover, our fabrication procedure based on ion beam etching (IBE) offers more flexibility compared to lift-off based structures, because the overall layer thickness of the NIM can be significantly increased, which allows to create multiple functional layers in a single processing step. 

\section{Experimental details}
The sample geometry for the hexagonal double fishnets is illustrated in figure $1$. The structure consists of a gold-silica-gold MIM stack $(60-60-60 nm)$ perforated with a hexagonal array of round holes on top of a glass substrate.

\begin{figure}[h]
\begin{center}
\includegraphics[width=8.7cm]{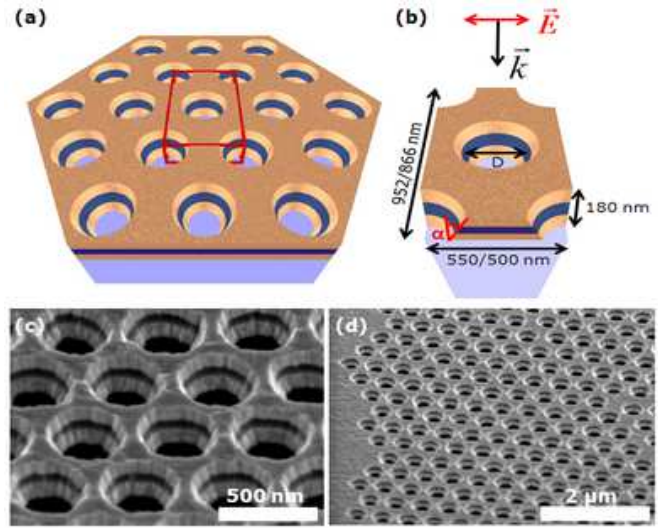}
\caption{Self-assembled hexagonal double fishnet sample structure. (a) Schematic overview of the structure. (b) Unit cell used in simulations. D denotes the diameter and $\alpha$ the $20\,^{\circ}$ sidewall angle of the holes. The MIM layer structure consists of $Au-SiO_2-Au$ $(60-60-60 nm)$. Unit cell dimensions are given for the self-assembly (larger) / e-beam (smaller) samples respectively. (c) Detailed scanning electron microscope (SEM) picture of the sample. (d) Overview SEM picture showing the edge of a perfectly ordered domain and some defects (right-hand side).}
\label{figure}
\end{center}
\end{figure}

\subsection{Sample fabrication}

The samples were fabricated by colloidal lithography using $550$ nm PS beads. The fabrication process for the self-assembled samples is outlined in figure $2$. The MIM layers are sputter deposited onto the glass substrate and subsequently covered with MET-2D \cite{Dow} e-beam resist by spincoating. The resist layer is covered with a $10$ nm gold protection layer by sputter deposition (panel a). On top of this layer stack, a hexagonal close-packed monolayer of PS beads is deposited by spincoating, which yields particle arrays up to $100$ x $100 \mu m^2$ that have a single lattice orientation and a small number of defects. In the next step, the PS bead positions are first fixed on the substrate by a short annealing step ($1$ minute at $100^{\circ}{\rm C}$ under nitrogen atmosphere) and subsequently the beads are shrunk using oxygen plasma etching. In this step the gold protection layer prevents etching of the MET-2D resist layer, while the duration of the etch allows to tune the final hole diameter. The PS bead array then serves as the template for the hole array by covering them by evaporation with a $10$ nm titanium layer (panel b) and performing a lift-off by dissolving the PS beads in toluene (panel c). The hole pattern is then transferred into the Au protection layer by IBE (panel d) and into the resist layer by inductively coupled oxygen plasma etching (panel e). As a final etching step, the hole pattern is transferred from the resist into the MIM by IBE, while the remaining resist is removed using an oxygen plasma etch (panel f). During the IBE step, a sidewall slope of about $20^{\circ}$ is introduced in the holes (fig 1.b).

\begin{figure}[h]
\begin{center}
\includegraphics[width=8.7cm]{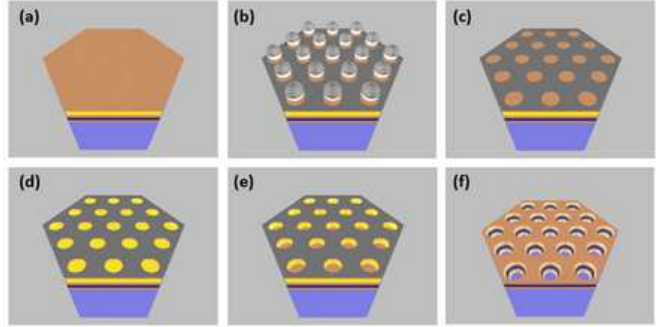}
\caption{Sample fabrication procedure. (a) Deposition of the layer stack: sputter deposition of $Au-SiO_2-Au$ MIM, spincoating of the MET-2D resist (yellow layer) and sputter deposition of Au protection layer. (b) Spincoating of PS beads, bead shrinking by oxygen plasma etching and evaporation of Ti hard mask. (c) lift-off by dissolving PS beads. (d) Ion beam etch (IBE) through gold protection layer. (e) Transfer pattern into MET-2D resist by inductively coupled oxygen plasma etching. (f) IBE of holes in MIM layer stack and removal of remaining resist. }
\label{figure}
\end{center}
\end{figure}

For reference purposes we also made samples by e-beam lithograpy. The colloidal lithography step is replaced by e-beam exposure and development of the MET-2D resist (there is no need for a sacrificial gold protection layer). The hole pattern is etched into the MIM stack immediately after development by IBE and the remaining resist is removed using oxygen plasma etching. 

\subsection{Optical characterization}

The optical response of our metamaterial samples was recorded using a home-built transmission and reflection measurement setup. A supercontinuum fiber laser (Fianium SC450-4) equipped with a Near-infrared acousto-optical tunable filter (AOTF) was used to illuminate the samples by a single wavelength. The incident beam was focussed onto the sample using an apochromatic 20X NIR (Newport N-20X-APO-IR) objective. The reflected light was collected through the same objective while the transmitted light was collected at the backside using a similar objective. The beam intensities of the transmitted and reflected light were recorded using a silicon photodiode. The transmission spectra were normalized with respect to a bare glass substrate and the reflection spectra were normalized with respect to a continuous gold film. 

\subsection{Simulations and effective parameters}

The electromagnetic response of our samples was modeled using finite-difference time domain (FDTD) simulations, using the commercial package Lumerical FDTD \cite{Lum}. The dielectric properties of the gold layers were taken from Johnson and Christy \cite{JandC}. A rectangular unit cell and a mesh of $2$ x $2$ x $2$ $nm^3$ was used (figure 1 a and b). The structure is modeled using periodic boundary conditions in the plane of the MIM and with perfectly matched layers that absorb all transmitted or reflected power at the top and bottom of the unit cell. A perpendicularly incident plane wave excitation was used with the polarization aligned parallel to the hole pattern (fig 1.b). The complex fields on the top and bottom surface of the MIM trilayer structure were averaged over the unit cell (homogenization) in order to extract the effective material parameters \cite{Smith}, taking into account the bianisotropy of the samples in the propagation direction. In that way, the complex effective parameters (refractive index n, impedance Z, permittivity $\epsilon$ and permeability $\mu$) and the FOM were calculated. In order to compensate for additional damping due to the nanostructure scale and interface roughness of the metamaterial, the complex part of the permittivity of gold was increased (up to 3 times the bulk value) with respect to the bulk material parameters, to allow comparison of the extracted effective parameters to earlier works \cite{Dolling1, Dolling2, Dolling3, Chettiar, Xiao}. 

\section{Results and discussion}

\begin{figure}[h]
\begin{center}
\includegraphics[width=8.7cm]{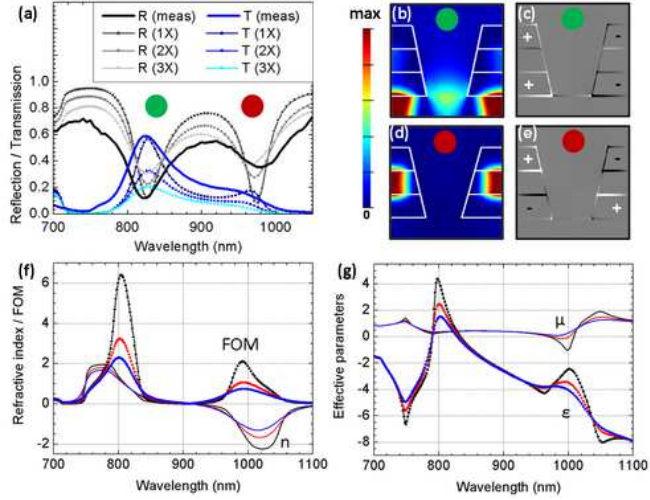}
\caption{Measurement and simulation results for a self-assembled hexagonal double fishnet. (a) Measured reflection (black full line) and transmission (blue full line) spectra and simulated (dotted lines) reflection and transmission spectra with increasing imaginary part of the permittivity in the gold layers. ($1,2,3X =$ number of times the bulk imaginary permittivity value). (b) Out-of-plane magnetic field intensity plot at 825 nm resonance. (c) Charge density plot at 825 nm resonance. (d) Out-of-plane magnetic field intensity plot at 980 nm resonance. (e) Charge density plot at 980 nm resonance. (f) Simulated real part of the refractive index (n) and figure of merit (FOM) for different values of damping in the gold layers ($black = 1X$, $red = 2X$ and $blue = 3X$). (g) Simulated effective parameters: real parts of permittivity and permeability for different values of damping in the gold layers ($black = 1X$, $red = 2X$ and $blue = 3X$).}
\label{figure}
\end{center}
\end{figure}

An overview of the optical response of the self-assembled hexagonal double fishnets with a hole diameter of 270 nm and a pitch of 550 nm is given in figure 3. The measured and simulated transmission and reflection data (panel a) show two pronounced resonances of the structure which exhibit good qualitative agreement. The corresponding magnetic field intensity plots (panels b and d) and charge density plots (panels c and e) provide good insight in the nature of these resonant modes. The magnetic field intensity plots show the out-of-plane field component in a cross section of the MIM layer stack along the electric field component of the incident plane wave. This perpendicular component is a good measure for the excitation of surface plasmons, and illustrates how the first resonant mode around 825 nm (panel b) is mainly confined at the interface between the bottom gold layer and the substrate, whereas the second resonant mode around 980 nm (panel d) is mainly confined on the top and bottom interfaces of the insulator layer. The corresponding charge density plots show that for the first mode (panel c) parallel displacement currents and for the second mode (panel e) anti-parallel displacement currents are excited in the top and bottom gold layers of the MIM cavity. The second mode clearly shows a strong magnetic resonance (panel d) where a magnetic dipole is excited in the oxide gap of the MIM cavity, which is out-of-phase with the magnetic field component of the incident plane wave, and gives rise to a strong decrease of the effective permeability $\mu$ of the metamaterial. In combination with the negative permittivity $\epsilon$ of the gold layers in this wavelength range, this results in a negative value for the refractive index. For this particular sample, the effective material parameters extracted from the simulations are summarized in panels f and g. We first carried out the extraction with the bulk properties of gold \cite{JandC} and then gradually increased the imaginary part of the permittivity up to three times the bulk value, in order to compensate for fabrication imperfections and interface roughness. When increasing the imaginary part of the permittivity of the gold layers, all resonances are damped (decreased amplitude) and broadened while maintaining their spectral position. The damping of the resonances is also reflected in the extracted effective parameters, which show a decrease in the resonance amplitude as the imaginary permittivity of the gold layers is increased. As for the refractive index, smaller absolute values are obtained for the real part, while the imaginary part is increased, resulting in a decrease of the FOM of the magnetic resonance (panel f). For the real parts of the permittivity and permeability, the amplitude of the resonance is decreased as the imaginary permittivity of the gold is increased (panel g). When comparing the measured spectra with simulation data, we clearly observe that for the first resonance, the bulk damping coefficients give us the best fit, which can be explained by the nature of this resonance. The mode is dominated by the bottom hole cavity and plasmons excited at the interface between the substrate and the bottom gold layer (panel b). Since the substrate roughness is substantially smaller than the roughness of the sputtered MIM layers, we don't expect much additional damping on top of the bulk material properties \cite{Nagpal}. For the magnetic resonance on the other hand, we clearly see more damping in the measurement than for the simulation with the bulk material parameters (1X in panel a). When increasing the imaginary part of the permittivity (2X and 3X) in the simulations, we clearly see the modes becoming less pronounced, which is most apparent in the transmission spectrum near the magnetic resonance.  The step-like behavior is observed in all the simulations, but the sharp edge that is observed for the bulk imaginary permittivity becomes less steep as the damping is increased. Therefore, the step-like behavior in the transmission spectrum is a good measure to determine the importance of damping, by fitting the measured spectrum to the simulated spectra with different values of the imaginary permittivity. In practice, we mainly look at a qualitative correspondence of the shape of the magnetic resonance to determine the most appropriate damping. Based on our experimental results for various hole sizes, we suggest that doubling the imaginary permittivity of the gold layers is sufficient to account for the increased damping due to surface roughness and the nanoscale dimensions of the metamaterial. Again, the observation of increased damping can be understood from the nature of the resonance. At the magnetic resonance, plasmons are excited on the top and bottom of the dielectric spacer, which are more prone to surface roughness on both of the interfaces, as these are created by sputter deposition \cite{Nagpal}.

\begin{figure}[h]
\begin{center}
\includegraphics[width=8.7cm]{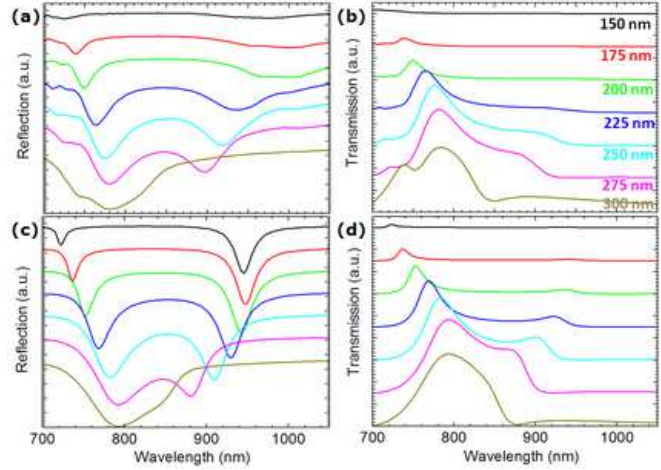}
\caption{Measured and simulated reflection and transmission measurements for e-beam lithography based hexagonal double fishnets with hole diameters ranging from 150 to 300 nm. (a) Measured reflection spectra. (b) Measured transmission spectra. (c) Simulated reflection spectra. (d) Simulated transmission spectra. }
\label{Figure}
\end{center}
\end{figure}

In order to obtain a more complete dataset and to confirm the nature of the observed resonances, we also fabricated samples by e-beam lithography for reference purposes.

Figure 4 shows an overview of the measured and simulated transmission and reflection spectra for a batch of hexagonal double fishnets with different hole diameters ranging from 150 nm to 300 nm with a pitch of 500 nm. The measured spectra show excellent agreement to the simulated spectra, in which the bulk material parameters were used. For all samples, the two main modes of the structure can be clearly observed, and show the expected wavelength shifts with respect to the hole diameter. The first resonance is observed between 750 nm to 800 nm, and shows a red-shift with increasing hole diameter. Around this resonance, plasmons are excited at the interface between the substrate and the bottom gold layer, giving rise to strong magnetic field confinement in the bottom hole cavity (fig 3.b). This mode clearly shows a red-shift of the resonance with increasing hole diameter, which is related to the cut-off of the hole transmission. As the hole size is decreased, the resulting decrease in transmission is smaller for short wavelengths as compared to larger wavelengths, due to the non-linear dependence of the transmission beyond the cut-off frequency of the hole waveguide \cite{Molen}. The second mode is observed between 900 to 950 nm and shows a blue-shift with increasing hole diameter. The magnetic field intensity plot (fig 3.d) clearly shows strong magnetic field intensity in the oxide gap of the MIM cavity in between the holes, due to the excitation of gap plasmons \cite{Mary} on the top and bottom interface of the oxide layer. As the hole diameter increases, the cavity length of the MIM in between the holes is reduced, which explains the observed blue shift of the resonance.

All samples shown in figure 4 exhibit negative values for the refractive index around the magnetic resonance. The strength of the magnetic resonance is different for the various hole sizes, and the strongest resonance is observed for a hole diameter of 250 nm. When using the optimized material parameters for the gold layers, with the imaginary part of the permittivity being twice the bulk value, we observe double negative behavior (both real parts of permeability and permittivity show negative values at the same wavelength) with the highest figure of merit of all the investigated samples. The resulting real part of the permeability reaches a minimum of -0.1 and a real part of the refractive index of -1.4, with a FOM of 0.95. For similar simulation conditions, all other samples exhibit single negative behavior (only the real part of permittivity is negative) with lower FOMs.

When we compare the measured spectra for self-assembly based samples (figure 3.a) and e-beam lithography based samples (figure 4 a and b), we observe that the spectra are very similar in nature, but of course there are some some differences. For both sample types, we observe broadening of the measured resonances with respect to the simulated spectra, which can be related to
fabrication imperfections and roughness of the different layers. As outlined before, this can be accounted for in simulations by increasing the damping in the gold layers. However, if we compare the self-assembly samples to the e-beam samples, we can also clearly observe that the broadening is more pronounced for self-assembled samples (Q-factor of about 15 for self-assembly samples versus 20 for e-beam samples), as a consequence of imperfections in the structure. These comprise line defects and vacancies (missing holes), but also variations in hole sizes due to small size variations in the PS beads and imperfect hole shapes due to the PS bead shrinking step. Although these imperfections are intrinsically present due to the limitations of the fabrication process, they clearly only have a minor influence on the material parameters of the overall metamaterial structure.

\section{Conclusions}

To summarize, we have shown the feasibility to use nanosphere lithography for the fabrication of large area metamaterials that exhibit a negative value of the refractive index in the near-infrared wavelength range. These samples were compared with samples fabricated by e-beam lithography, and show very good agreement. Depending on the hole diameter, the metamaterials can be tuned to be single- or double-negative in nature. Moreover, we have shown that the damping of the resonances due to the nanoscale structure and fabrication imperfections seems to be smaller compared to earlier works, and should in our case only be taken twice as large as the bulk material parameters for gold in order to allow extraction of effective material parameters.

\ack
K.L. acknowledges financial support from I.W.T. (Flanders), N.V. acknowledges support from the Methusalem funding by the Flemish government and P.V.D. acknowledges financial support from F.W.O. (Flanders). We thank Jos Moonens for e-beam assistance
and Erwin Vandenplas, Peter Vicca and Steve Smout for processing assistance.


\section*{References}
\numrefs{1}
\bibitem{Veselago} Veselago V G 1968 {\it Sov. Phys. USPEKHI} {\bf 10} 509
\bibitem{Pendry} Pendry J B 2000 {\it Phys. Rev. Lett.} {\bf 85} 3966
\bibitem{Dolling1} Dolling G, Enkrich C, Wegener M, Soukoulis C M and Linden S 2006 {\it Opt. Lett.} {\bf 31} 1800
\bibitem{Dolling2} Dolling G, Enkrich C, Wegener M, Soukoulis C M and Linden S 2006 {\it Science} {\bf 312} 892
\bibitem{Dolling3} Dolling G, Wegener M, Soukoulis C M and Linden S 2007 {\it Opt. Lett.} {\bf 32} 53
\bibitem{Chettiar} Chettiar U K, Kildishev A V, Yuan H-K, Cai W, Xiao S, Drachev V P and Shalaev V M 2007 {\it Opt. Lett.} {\bf 32} 1671
\bibitem{Xiao} Xiao S, Chettiar U K, Kildishev A V, Drachev V P and Shalaev V M 2009 {\it Opt. Lett.} {\bf 34} 3478
\bibitem{Valentine} Valentine J, Zhang S, Zentgraf T, Ulin-Avila E, Genov D A, Bartal G and Zhang X 2008 {\it Nature} {\bf 455} 376
\bibitem{Mary} Mary A, Rodrigo S G, Garcia-Vidal F J and Martin-Moreno L 2008 {\it Phys. Rev. Lett.} {\bf 101} 103902
\bibitem{Zhou} Zhou J, Koschny T, Kafesaki M, Economou E N, Pendry J B and Soukoulis C M 2005 {\it Phys. Rev. Lett.} {\bf 95} 223902
\bibitem{Klein} Klein M W, Enkrich C, Wegener M, Soukoulis C M and Linden S 2006 {\it Opt. Lett.} {\bf 31} 1259
\bibitem{Dow} Dow Corning Corporation, Corporate Center, PO Box 994, MIDLAND MI 48686-0994, United States. www.dowcorning.com
\bibitem{Lum} Lumerical FDTD, Lumerical solutions inc., Suite 201 - 1290 Homer Street Vancouver, British Columbia, Canada. www.lumerical.com/fdtd
\bibitem{JandC} Johnson P B and Christy R W 1972 {\it Phys. Rev. B} {\bf 6} 4370
\bibitem{Smith} Smith D R, Vier D C, Koschny Th and Soukoulis C M 2005 {\it Phys. Rev. E} {\bf 71} 036617
\bibitem{Nagpal} Nagpal P, Lindquist N C, Oh S-H and Norris D J 2009 {\it Science} {\bf 325} 594
\bibitem{Molen} van der Molen K L, Segerink F B, van Hulst N F and Kuipers L 2004 {\it Appl. Phys. Lett.} {\bf 85} 3416
\endnumrefs
\end{document}